\documentclass[twocolumn]{aastex6}
\received{Sep 19, 2017}
\revised{XXXX}
\accepted{XXXX}
\shorttitle{Calvera}
\shortauthors{Li et al.}
\begin{document}
\title{Calvera: A low-mass strangeon star torqued by debris disk?}

\author{Yunyang Li}
\affiliation{School of Physics and State Key Laboratory of Nuclear Physics and Technology, Peking University, Beijing 100871, China}
\email{liyunyang@pku.edu.cn, liyunyang95@gmail.com}

\author{Weiyang Wang}
\affiliation{School of Physics and State Key Laboratory of Nuclear Physics and Technology, Peking University, Beijing 100871, China}
\affiliation{Key Laboratory of Computational Astrophysics, National Astronomical Observatories, CAS, Beijing 100012, China}
\affiliation{School of Astronomy and Space Sciences, University of Chinese Academy of Sciences, Beijing 100049, China}

\author{Mingyu Ge}
\affiliation{Key Laboratory for Particle Astrophysics, Institute of High Energy Physics, Chinese Academy of Sciences, Beijing 100049, China}
\author{Xiongwei Liu}
\affiliation{School of Physics and Space Science, China West Normal University, Nanchong 637002, China}
\author{Hao Tong}
\affiliation{School of Physics and Electronic Engineering, Guangzhou University, Guangzhou 510006, China}
\author{Renxin Xu}
\affiliation{School of Physics and State Key Laboratory of Nuclear Physics and Technology, Peking University, Beijing 100871, China}
\affiliation{Kavli Institute for Astronomy and Astrophysics, Peking University, Beijing 100871, China}

\begin{abstract}
Calvera is a $59\,\mathrm{ms}$ isolated pulsar, being unique due to its non-detection in radio, optical and gamma-rays but the purely thermal emission in soft X-rays. 
It is suggested that Calvera could be an ordinary middle-aged pulsar with significant magnetospheric activity at a large distance \citep{2016ApJ...831..112S}.
Alternatively, it is proposed in this paper that Calvera is a low-mass strangeon star with inactive magnetosphere (dead).
In this scenario, we jointly fit the spectra obtained by the
{\it XMM-Newton} Observatory and the {\it Chandra} X-ray Observatory with the strangeon star atmosphere model. The spectral model is successful in explaining the radiation properties of Calvera and X-ray Dim Isolated Neutron Stars, both showing similar observation features.
Within the dead pulsar picture, Calvera might be of high temperature at $0.67\,\mathrm{keV}$, possessing a small stellar radius $R\la4\,\mathrm{km}$ and a presumably small magnetic field $B\la10^{11}\,\mathrm{G}$ and  and is probably braked by the fall-back disk accretion.
Future advanced facilities may provide unique opportunities to know the real nature of Calvera.

\end{abstract}

\keywords{pulsars: individual (Calvera, 1RXS J141256.0$+$792204) -- stars:neutron -- accretion}



\section{Introduction} \label{sec:intro}

The {\it ROSAT} All-Sky Survey discovered a high galactic latitude ($b=37\arcdeg$) compact object, 1RXS J141256.0+792204 \citep{2008ApJ...672.1137R}, which was then identified as an isolated neutron star (INS, hereafter ``NS'' refers to all kinds of pulsar-like compact objects) candidate.
The fact that this INS is discovered after the seven radio-quiet and thermally emitting X-ray dim isolated neutron star (XDINS), the Magnificent Seven \citep[see][for reviews]{2007Ap&SS.308..181H,2008AIPC..983..331K}, leads it to be nicknamed as ``Calvera''.
Calvera, in particular, is a puzzling source that has some confusion in the classification among the neutron star family.

\begin{figure}
\epsscale{1.3}
\plotone{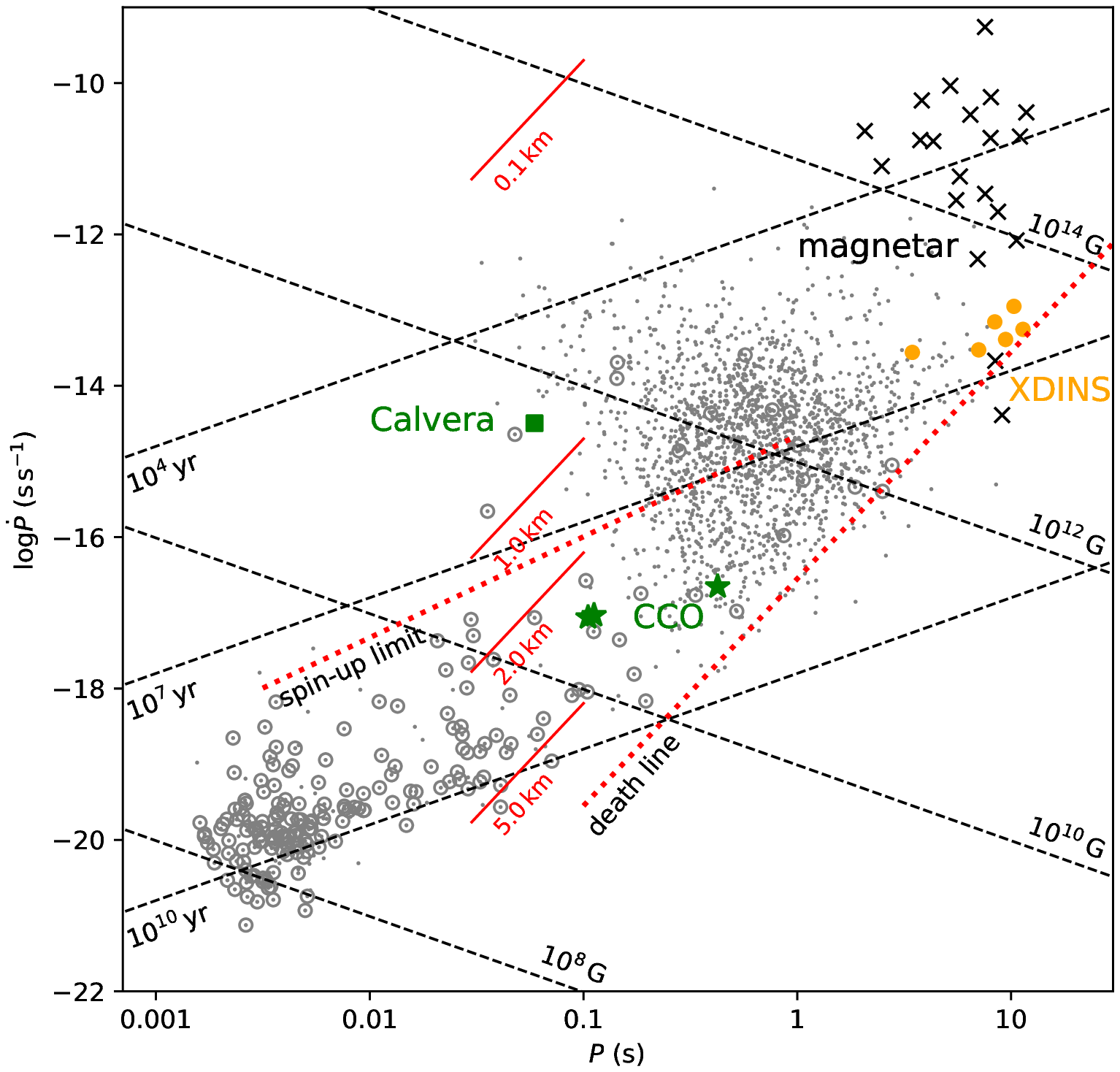}
\caption{The $P-\dot{P}$ diagram for radio pulsars (gray dots), binary pulsars (circlets), magnetars (crosses), XDINSs (orange dots), CCOs (green stars) and Calvera (green square). The pulsar population data are from  \href{http://www.atnf.csiro.au/research/pulsar/psrcat}{ATNF Pulsar Catalogue} \citep{2005AJ....129.1993M}. The spin data of the three CCOs are from \cite{2010ApJ...709..436H} and \cite{2013ApJ...765...58G}. The spin-up limit is shown as the upper red dotted line \citep[$P=1.9\,(B/10^9\,\mathrm{G})^{6/7}\,\mathrm{ms}$,][]{1987IAUS..125..393V}. The death line for a typical $R=10\,\mathrm{km}$ neutron star is indicated by the lower red dotted line \citep[$BP^{-2}=1.7\times 10^{11}\,\mathrm{G\,s^{-2}}$,][]{1992A&A...254..198B}. The death lines for low-mass strangeon star with two times the nuclear density and different stellar radii are denoted as red solid lines, on the assumption that a pulsar is torqued by magnetospheric activity. \label{ppdot}}
\end{figure}

Calvera exhibits X-ray pulsations with period $P = 59\,\mathrm{ms}$ \citep{2011MNRAS.410.2428Z} and spin-down rate $\dot{P} = 3.2\times10^{-15}\,\mathrm{s\,s^{-1}}$ \citep{2013ApJ...778..120H,2015ApJ...812...61H}, making its location in the $P-\dot{P}$ diagram (Figure \ref{ppdot}) far from the Magnificent Seven class, which are slowly rotating $(P\sim3-11\,\mathrm{s})$ NSs. It is also speculated that Calvera might be a candidate of the central compact object \citep[CCO,][]{,2008ApJ...672.1137R,2011MNRAS.410.2428Z,2013ApJ...765...58G}. 
However, Calvera presents a larger dipole magnetic field \citep{2009ApJ...705..391S} and there is still no conclusive evidence for the host supernova remnant \citep[][]{2011MNRAS.410.2428Z}. 
Alternatively, there are suggestions that the magnetic field ($\sim10^{12}\,\mathrm{G}$) of CCO is buried by prompt fall-back of a small amount of supernova ejecta \citep{2011MNRAS.414.2567H,2012MNRAS.425.2487V,2013ApJ...770..106B}; therefore, Calvera could be a descendant of the CCO reemerging the magnetic field \citep{2013ApJ...778..120H}.

It is odd that deep search failed to detect the radio emission from this source \citep{2007A&A...476..331H,2011MNRAS.410.2428Z}.
The non-detection of radio emission from Calvera can not be simply attributed to the unfavored beaming effect since emission features are nor found in gamma-rays \citep{2011ApJ...736L...3H,2013ApJ...778..120H} which commonly correspond to a larger beaming angle.
This can be explained by a distant location of Calvera \citep[e.g.,$1.5-5\,\mathrm{kpc}$,][]{2016ApJ...831..112S}, but that would place it high above the Galactic disk and cause problem for its birth place, given its rather small proper motion \citep[$69\pm26\,\mathrm{mas\,yr^{-1}}$,][]{2015ApJ...812...61H}.
Moreover, attempts also failed in searching for non-thermal emission feature in the soft X-ray band \citep{2011MNRAS.410.2428Z,2013ApJ...778..120H}.
All these observational facts could contain hints for an inactive-magnetosphere (i.e., dead) scenario which will be discussed in this work.

The dead-pulsar-scenario is hardly understood in the framework of NS due to its high position above the NS death line (Figure \ref{ppdot}). Spectral fits for Calvera with neutron star atmosphere model result in small emission-radius-to-distance ratio, which forces \cite{2016ApJ...831..112S} to conclude a large distance. Alternatively, we explore the possibility that Calvera is a small radius strangeon star.
``Strangeon'' \citep{2017JPhCS.861a2027X}, previously known as ``strange quark-cluster'' \citep{2003ApJ...596L..59X} is a prospective candidate for the pulsar constituent and have been successfully applied to solve problems including glitches \citep{2014MNRAS.443.2705Z}; high mass NS \citep{2009MNRAS.398L..31L,2011RAA....11..687L}; ultra low-mass and small radius NS \citep{2015ApJ...798...56L}. A radiative model of the strangeon star atmosphere \citep[SSA,][]{2017ApJ...837...81W,2017arXiv170503763W} is also developed to solve the optical/ultra-violet(UV) excess problem \citep{1997Natur.389..358W,2001A&A...378..986V,2011ApJ...736..117K} and the Rayleigh-Jeans deviation problem \citep{2011ApJ...736..117K} of XDINSs. 
The luminosity of strangeon star is maintained by accretion \citep{2017ApJ...837...81W} that would also exert a braking torque accounting for the spin-down rate.
It is proposed here that Calvera is a low-mass strangeon star with an inactive magnetosphere and probably braked by the accretion flow. In this picture, Calvera and XDINSs, having similar radiative properties, can be understood as strangeon stars at different stages of the evolution.

We introduce the data reduction procedure and spectra modeling in Section \ref{sec:data} and \ref{sec: spec_model}, respectively. In Section \ref{sec:deathline} we constrain the parameters of Calvera as an isolated dead-pulsar while in Section \ref{sec:accretion} we re-consider this issue by taking into account the accretion effects.  We discuss the nature of Calvera in Section \ref{sec:disc}. Summary and future possible observations in constraining the nature of Calvera are presented in Section \ref{sec:summary}.

\section{Data Reduction} \label{sec:data}
Since the first detection by {\it ROSAT} \citep{1999A&A...349..389V}, X-ray observations on Calvera have been made several times by {\it Swift} \citep{2008ApJ...672.1137R}, {\it XMM-Newton} \citep{2011MNRAS.410.2428Z} and {\it Chandra} \citep{2009ApJ...705..391S,2013ApJ...778..120H,2015ApJ...812...61H}.
Despite numerous attempts in modeling the X-ray spectra, the nature of Calvera still remains open. In this work, we make a joint analysis \citep[also see,][]{2016ApJ...831..112S} for the spectral data obtained by {\it XMM-Newton} and {\it Chandra} to further investigate the properties of Calvera. {\it Swift} data are omitted due to its limited counting statistics \citep{2008ApJ...672.1137R}.
\subsection{\it Chandra}
We retrieved the {\it Chandra} Advanced Camera for Imaging and Spectroscopy (ACIS) data from the public archive.
Among which, one \citep[obs.ID 9141,][]{2009ApJ...705..391S} operated in the VFAINT mode and two \citep[obs.ID 13788,15613,][]{2013ApJ...778..120H} in the continuous-clocking (CC) mode.
The data reduction and analysis were performed with the {\it Chandra} Interactive Analysis of Observation \citep[CIAO, version 4.9,][]{2006SPIE.6270E..1VF} with calibration database (CALDB 4.7.4).
For obs.ID 9141, we extracted $3599$ source photons from a circle centered on the target with radius $4.16\arcsec$ and $92$ background counts from the annulus surrounding the source region with an outer radius $8.32\arcsec$.
For data obtained in the CC mode, source counts were extracted from a $5$-column-box (15 pixels) centered on the target and the background counts from a $5$-column-box away from the source.
The two CC mode observations were weighed by the exposure time and combined together.
All {\it Chandra} spectra were grouped with a minimum of $25$ counts per bin.
We used the {\it Chandra} ``soft band'' $0.5-2.0\,\mathrm{keV}$ for modeling.
\subsection{XMM-Newton}
The data reduction for {\it XMM-Newton} were performed with Science Analysis System \citep[SAS, version 16.0.0,][]{2014ascl.soft04004S}. We utilized the European Photon Imaging Camera(EPIC)-pn data of the {\it XMM-Newton} observations \citep[obs.ID: 0601180101,0601180201,][]{2011MNRAS.410.2428Z}.
Data from the two EPIC-MOS cameras were not used in our analysis due to its smaller effective area at soft X-ray band \citep{2001A&A...365L..27T}.
All observations were obtained in Small Window (SW) mode with the thin filter.
Good time intervals were chosen according to the light curves at $0.1-5\,\mathrm{keV}$ band.
The source photons were extracted from the circular region with radii $15\arcsec$ and the background from the adjacent source free region of same size.
For spectral analysis, we selected single pixel events (PATTERN=0) and excluded bad CCD pixels and columns (FLAG=0).
The {\it XMM-Newton} spectra were grouped with at least 30 counts per bin; events within the $0.1-3.0\,\mathrm{keV}$ range were adopted for modeling.

The information of the data used are summarized in Table \ref{tab:data} for reference. All spectra modeling were performed with XSPEC version 12.9.1 \citep{1996ASPC..101...17A}.

\begin{deluxetable*}{llcccccc}
\tabletypesize{\scriptsize}

\tablecaption{Summary of the X-ray data \label{tab:data}}
\tablewidth{0pt}
\tablehead{
\multicolumn{2}{c}{Data} & \colhead{Instrument/mode} & \colhead{Counts}  & \colhead{Exposure Time} & \colhead{Start Date} & \colhead{End Date} & \colhead{}\\
\multicolumn{2}{c}{    } & \colhead{    } & \colhead{      }  & \colhead{($\mathrm{ks}$)} & \colhead{} & \colhead{} & \colhead{}
}
\startdata
              & 9141 & ACIS-S(VF) & $3599$ & $26.43$ & 2008-04-08 03:42:08   & 2008-04-08 12:13:24 \\
{\it Chandra} & 13788 & ACIS-S(CC) & $2356$ & $19.68$ & 2013-02-12 19:28:07 &  2013-02-13 01:24:58  \\
              & 15613 & ACIS-S(CC) & $2155$ & $17.09$ & 2013-02-18 02:52:52 &  2013-02-18 08:03:38  \\
\hline
{\it XMM-Newton} & 0601180101 & EPIC-pn(SW) & $8921$ & $13.94$ & 2009-08-31 07:07:52 & 2009-08-31 15:08:42 \\
                 & 0601180201 & EPIC-pn(SW) &$11411$ & $19.48$ & 2009-10-10 04:08:42 & 2009-10-10 12:26:09 \\
\enddata
% \tablecomments{}
\end{deluxetable*}

\section{Spectral modeling}\label{sec: spec_model}
It is suggested that pulsars could be strangeon stars \citep{2017JPhCS.861a2027X}.
A strangeon star can be thought as a 3-flavored gigantic nucleus, and strangeons (coined by combining ``strange nucleons'') are its constituent as an analogy of nucleons which are the constituent of a normal (micro) nucleus.
The radiative model of the strangeon star atmosphere was put forth by \citep{2017ApJ...837...81W}. For an isolated strangeon star, normal matter (i.e., composed by u, d quarks) accreted onto the stellar surface can not be converted to strangeons (i.e., strangeonization) instantly because the collision timescale $\tau_\mathrm{col}\sim10^{-22}-10^{-20}\,\mathrm{s}$ is far smaller than that of the weak interaction $\tau_\mathrm{weak}\sim10^{-7}\,\mathrm{s}$. Therefore the unconverted matter would be re-bounced and form a thermally emissive atmosphere. The flux can be described as
\begin{equation}
F_\nu^\infty \propto B_\nu(1-e^{-2\tau_\infty(\nu)}),
\end{equation}
where $F_\nu$ is the flux at frequency $\nu$ and $B_\nu$ is the blackbody spectrum. $\tau_\infty(\nu)$ is the observed optical depth, with the factor $2$ accounting for the surface reflection \citep{2017arXiv170503763W}, and can be expressed as
\begin{equation}
\tau_\infty(\nu) = 3.92\times10^{-45}\frac{n^2_{\mathrm{i}0}(kT_\mathrm{i})_\mathrm{keV}}{(h\nu)^{3.5}_\mathrm{keV}R_\mathrm{km}}(1-e^{-\frac{h_\nu}{kT_\mathrm{e}}}),
\end{equation}
where $n_{\mathrm{i}0}$ is the ion density at bottom, $T_\mathrm{i}$ and $T_\mathrm{e}$ are the ion and electron temperatures, respectively, and $R$ is the stellar radius. 
We use the notation $y={n^2_\mathrm{i0}(kT_\mathrm{i})_\mathrm{keV}}/R_\mathrm{km}\sim10^{42}\,\mathrm{keV\,km^{-1}\,cm^{-6}}$ for these degenerate parameters which is different from the one used in \cite{2017ApJ...837...81W,2017arXiv170503763W} by $1/R$ (the SSA model with this new definition is now uploaded to \href{https://heasarc.gsfc.nasa.gov/xanadu/xspec/models/ssa.html}{Xspec} for public use).
At lower energies (i.e., optical/UV bands), the optical depths are high and the radiation behaves like a blackbody.
Whereas for soft X-rays, the typical optical depths are small and the flux can be approximated by $F_\nu = 2\tau(\nu)B_\nu$, which is lower than a pure blackbody spectrum.
Therefore, extrapolating the blackbody spectrum obtained in the X-ray band will meet the optical/UV excess problem \citep{2011ApJ...736..117K}.
The low optical depths in X-rays have two consequences: 
(1) Parameter $y$ is partially degenerate with the normalization $(R_\mathrm{km}/d_{10\,\mathrm{kpc}})^2$ and the two cannot be determined simultaneously without the knowledge of the optical/UV flux. 
(2) Since the optical/UV excess is expected from a simple extrapolation of the blackbody spectrum in the X-rays, we can use the extrapolation of the pure blackbody fit for X-rays at optical band as a lower limit for the normalization. 
On the other hand, optical upper limit were obtained by {\it Gemini}-North \citep[g band,][]{2008ApJ...672.1137R} and Gran Telescopio Canarias \citep[GTC, $g^\prime,r^\prime$ bands,][]{2016ApJ...831..112S}, which give the upper limit of the normalization.

\begin{figure}
\epsscale{1.3}
\plotone{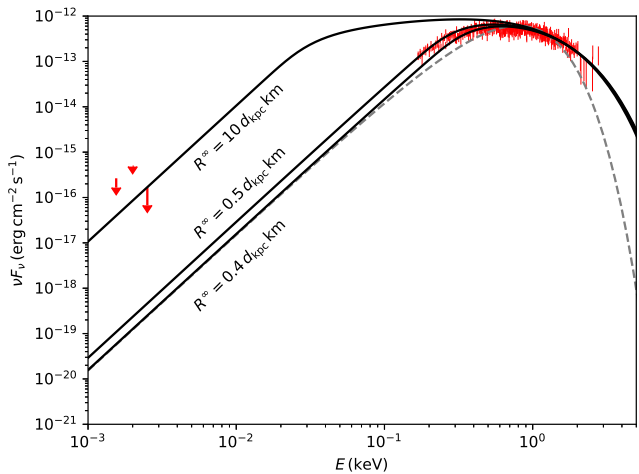}
\caption{Combined data (red bars) and the fitting results. The red triangle and arrows are the optical flux upper-limit given by {\it Gemini}-North\citep{2008ApJ...672.1137R} and GTC \citep{2016ApJ...831..112S}. The gray dashed line is the one-component blackbody fitting. The solid lines are the best-fits for model F, M1 and M2. Galactic absorption is not shown.  \label{joint_fitting}}
\end{figure}

The phase-averaged spectral analysis was performed simultaneously for data obtained with different detectors or at different times, allowing only the parameter $y$ to vary independently to account for possible cross-calibration uncertainties.
The fit was conducted with blackbody model (BB) and SSA model with fixed $N_\mathrm{H}$ \citep[F, with $N_\mathrm{H}$ fixed to the Galactic value,][]{2005A&A...440..775K} or $R^\infty$ (M1, M2).
The upper limit of the normalization (M1) was chosen such that the extrapolated spectrum meet the upper limit at GTC $g^\prime$ band.
The lower limit (M2) was chosen such that the extrapolated spectrum converges with the blackbody fit at lower energies (e.g., $\sim \mathrm{eV}$). Results are listed in Table \ref{tab:fit_result} and illustrated in Figure \ref{joint_fitting}.

As is shown in \cite{2009ApJ...705..391S}, \cite{2011MNRAS.410.2428Z} and \cite{2013ApJ...778..120H}, single thermal spectra, either blackbody or pure hydrogen atmospheric model (NSA), can not provide decent fit, and two thermal components is required. The first joint fit for all available data is performed by \cite{2016ApJ...831..112S}, who use a single component magnetized hydrogen atmosphere model to account for the inhomogeneities of the stellar surface. Assuming a magnetic field $B=10^{12}\,\mathrm{G}$, they obtain good fits in spite of the viewing geometry. These results are broadly consistent in the sense that $T^\infty\sim 200\,\mathrm{eV}$ for blackbody models and $T^\infty\la 100\,\mathrm{eV}$ for NSA models and emission area $R^\infty/d=2-4\,\mathrm{km\,kpc^{-1}}$\citep{2008ApJ...672.1137R,2009ApJ...705..391S,2011MNRAS.410.2428Z,2013ApJ...778..120H,2016ApJ...831..112S}. In our SSA model, we obtain a higher temperature at $\ga0.6\,\mathrm{keV}$ and rather unconstrained radius $R^\infty/d=0.4-10\,\mathrm{km\,kpc^{-1}}$. \cite{2011MNRAS.410.2428Z} report that two-thermal models result in $N_\mathrm{H}$ larger than the Galactic value while we find that acceptable $N_\mathrm{H}$ values can be achieved assuming smaller $R^\infty/d$ (F and M2).
The small stellar radii in the magnetized NSA model lead to the conclusion of a large distance \citep[$1.5-5\,\mathrm{kpc}$,][]{2016ApJ...831..112S}, however this is not a problem for the SSA model since strangeon star can be a few kilometers in radius.

Absorption features about $0.6-0.7\,\mathrm{keV}$ are also reported as lines \citep{2009ApJ...705..391S,2011MNRAS.410.2428Z,2016ApJ...831..112S} or edges \citep{2011MNRAS.410.2428Z}. 
We conducted similar fitting procedure and find that absorption edges or Gaussian absorption lines equally improve the fit for model F, M1 and M2. 
The results for the SSA model multiplied by a {\tt gabs} model are list in Table \ref{tab:fit_result}.
The additional absorption line is found at $0.73\pm0.03\,\mathrm{keV}$, consistent with \cite{2016ApJ...831..112S}.
The presence of absorption line is often attributed to the magnetic field. 
In this case, the absorption line might indicate $B=6\times10^{10}\,\mathrm{G}$ assuming electron cyclotron of neutron star, or $B=1.2\times10^{11}\,\mathrm{G}$ for stangeon star hydrocyclotron \citep{2012PhRvD..85b3008X}.

\begin{deluxetable*}{cccccccccc}
\tabletypesize{\scriptsize}
\tablecaption{Summary of spectral modeling for Calvera \label{tab:fit_result}}
\tablewidth{0pt}
\tablehead{
\colhead{Model}\tablenotemark{a} & \colhead{$N_\mathrm{H}$} & \colhead{$kT_e$} & \colhead{$y$} &\colhead{$R^\mathrm{\infty}$}  & \colhead{$E$} & \colhead{$\tau$\tablenotemark{b}}  & \colhead{EW}  &\colhead{$F_\mathrm{X}$($0.3-10\,\mathrm{keV}$)}            & \colhead{$\chi_\mathrm{\nu}^2/${d.o.f}}\\
& \colhead{($10^{20}\,\mathrm{cm^{-2}}$)} & \colhead{($\mathrm{keV}$)} & \colhead{($10^{42}\,\mathrm{keV\,km^{-1}\,cm^{-6}}$)}  &\colhead{($d_{\mathrm{kpc}}\,\mathrm{km}$)} & \colhead{($\mathrm{keV}$)} & \colhead{} & \colhead{($\mathrm{eV}$)}& \colhead{$10^{-13}\,\mathrm{erg\,cm^{-2}\,s^{-1}}$} & \colhead{  }}
\startdata
BB\tablenotemark{c} &  $0$  & $0.2$ & $-$ & $0.6$ & - & - & - & $6.2$  & $1.97/482$\\
\hline
F  & $2.7$       & $0.67\pm0.02$ & $5.6$ & $0.51\pm0.03$ & - & - & - & $8.4$  & $1.06/482$\\
M1 & $5.0\pm0.4$ & $0.64\pm0.02$ & $0.02$ & $10$          & - & - & - & $10.0$ & $1.04/482$\\
M2 & $1.3\pm0.2$ & $0.67\pm0.02$ & $10.1$ & $0.37$        & - & - & - & $7.7$  & $1.09/482$\\
% \hline
% F  & $2.7$       & $0.61\pm0.03$ & $17.1$ & $0.46\pm0.02$ & $0.63\pm0.02$ & $0.29\pm0.09$ & - & $8.5$  & $1.00/480$\\
% M1 & $5.7\pm0.5$ & $0.61\pm0.02$ & $0.04$ & $10$          & $0.63\pm0.03$ & $0.19\pm0.08$ & - & $10.7$ & $1.01/480$\\
% M2 & $1.4\pm0.2$ & $0.61\pm0.03$ & $27.0$ & $0.37$        & $0.61\pm0.02$ & $0.34\pm0.10$ & - & $7.8$  & $1.01/480$\\
\hline
F  & $2.7$       & $0.62\pm0.03$ & $7.6$ & $0.48^{+0.02}_{-0.01}$ & $0.72\pm0.03$ & $0.19\pm0.02$ & $38^{+27}_{-15}$ & $8.5$  & $1.00/479$\\
M1 & $5.5\pm0.4$ & $0.61\pm0.03$ & $0.02$ & $10$          & $0.73\pm0.03$ & $0.17\pm0.02$ & $25^{+14}_{-12}$ & $10.5$ & $1.00/479$\\
M2 & $1.2\pm0.2$ & $0.62\pm0.03$ & $12.5$ & $0.37$        & $0.72\pm0.02$ & $0.21\pm0.02$ & $44^{+24}_{-16}$ & $7.7$  & $1.02/479$\\
\enddata
\tablenotetext{a}{BB: blackbody fit. F:$N_\mathrm{H}$ fixed to the Galactic value; M1: $R^\infty$ fixed to the maximum value to meet the optical upper limit; M2:  $R^\infty$ fixed to the minimum value to meet the blackbody fit.}
\tablenotetext{b}{Optical depth at the absorption line center.}
\tablenotetext{c}{Errors not shown due to poor fit.}
% \tablecomments{}
\end{deluxetable*}

\section{Torqued by magnetospheric activity?} \label{sec:deathline}
In the vacuum gap model for radio emission of neutron stars, \cite{1975ApJ...196...51R} propose the idea of death line, below which the electric potential of the gap region is too low ($<\Phi_c= 10^{12}\mathrm{\, V}$) to generate electron-positron pairs for curvature radiation.
The maximum potential drop ($\Phi_\mathrm{m}$) above the surface of a neutron star is
\begin{equation}
\Phi_\mathrm{m} = \frac{2\pi^2}{c^2P^2}BR^3, 
\end{equation}
which yields the death line 
\begin{equation}
R_\mathrm{max,km}^3=1.52\times10^{2}B_{12}^{-1}P_\mathrm{s}^2, \label{eq:deathline}
\end{equation}
on the premise that gap sparking could occur if $\Phi_\mathrm{m}>10^{12}\,\mathrm{V}$.
Assuming that magnetic dipole radiation accounts for the spin-down, we plot the death lines on the $P-\dot{P}$ diagram (Figure \ref{ppdot}) for typical neutron star radius $R = 10\,\mathrm{km}$ (red dotted line) and smaller strangeon star radii (red solid lines). 
Note that strangeon star with smaller radius (i.e., smaller momentum inertia) would exhibit larger magnetic field than that indicated by the dashed lines in Figure \ref{ppdot}.  
To meet the criterion that Calvera is dead, the upper limit for the stellar radius is $0.66\,\mathrm{km}$, which yields a stellar  mass $M=3\times10^{-4}\,M_\odot$. 
Therefore, in the context of dead pulsar, Calvera can be interpreted as a low-mass strangeon star.
However, these values are extreme even for strangeon stars. 
This problem could be alleviated if alternative mechanisms contribute to the spin-down. 
In our model (see Section \ref{sec:accretion}), the X-ray emission is maintained by the accretion (with a rate $\dot{M}_\mathrm{X}=1.2\times10^{12}d_\mathrm{kpc}^2\,\mathrm{g\,s^{-1}}$ on to the surface) 
which could also provide a torque braking the rapid-rotating low-mass star and accounting for the observed $\dot P$.

\section{Torqued by fall-back disk?}\label{sec:accretion}
The detailed mechanism of the accretion has significant impact on the long-term evolution and leads to distinct observational consequences. Two kinds of accretion source are discussed in literature.

Interstellar medium (ISM) accretion is first proposed by \cite{1970ApL.....6..179O} to understand the X-ray luminosity of INS \citep[][also see \citealt{2000PASP..112..297T} for reviews]{1991A&A...241..107T,1993ApJ...403..690B}. 
In this scenario, old INS traveling slowly through dense ISM accrete efficiently and exhibit less luminous ($\la10^{31}\,\mathrm{erg\,s^{-1}}$) thermal soft X-ray spectra. 
This model may explain the optical excess from the X-ray extrapolation \citep{2000ApJ...537..387Z} that is supported by succedent observations \citep[e.g.,][]{2011ApJ...736..117K}.

However, the relatively high proper motion and low ambient ISM density of XDINS cast doubts on the ISM accretion picture. 
Alternatively, it is reasonable that not all matter are expelled in the supernova explosion \citep{1971ApJ...163..221C,1989ApJ...346..847C} and a fractional infalling material may form a fall-back disk. 
The propeller or accretion torques of the disk can explain the high spin-down rate of neutron stars \citep{2001ApJ...554.1245A,2000ApJ...534..373C,2009ApJ...702.1309E,2014MNRAS.444.1559E,2016MNRAS.457.4114B,2017MNRAS.470.1253E} including XDINSs, the anomalous X-ray pulsars and soft gamma-ray repeaters \citep[AXPs and SGRs, see][for reviews]{2014ApJS..212....6O,2017ARA&A..55..261K}, the latter are otherwise interpreted as magnetars \citep{1995MNRAS.275..255T}. 
Especially, in the picture of \cite{2001ApJ...554.1245A}, XDINSs, AXPs and SGRs which populate similar region in the $P-\dot{P}$ diagram can be unified by the asymptotic propeller/accretion mechanism with alternative pathways.
In this scenario, the X-ray luminosity of AXP/SGR is caused by accretion \citep{2016MNRAS.457.4114B}
while that of XDINS is produced by energy dissipation in the neutron star \citep{2001ApJ...554.1245A,2007Ap&SS.308..133A} or by intrinsic cooling \citep{2014MNRAS.444.1559E}, i.e., accretion onto the stellar surface is not necessarily assumed in the propeller phase.
However, matter inflows are observed in simulations of ISM accretion propeller \citep{2003ApJ...588..400R} as well as disk accretion propeller \citep{2017arXiv170408336R} and the portion of the accreting matter may be sufficient to maintain the SSA radiation.

Nevertheless, the NS-disk system is not expected to reside in vacuum. 
Therefore, we propose here that ISM regulates the debris/fall-back disk accretion as a supplement and can be the dominant accretion source in the late phase when the fall-back material depletes. In this {\em ISM-fed debris disk accretion} (IFDA) picture, we expect $\dot{M}_\mathrm{A}>\dot{M}_\mathrm{X}>\dot{M}_\mathrm{B}$ when the fall-back system forms, 
where $\dot{M}_\mathrm{B}$ is the ISM accretion rate at the Bondi radius \citep{1952MNRAS.112..195B} and ${\dot M}_{\rm A}$ is the accretion rate at the Alfv\'en radius \citep{1979ApJ...234..296G}. Note that $\dot{M}_\mathrm{A}$ decreases as the disk loses its mass gradually through accretion and propeller wind. In the late phase of the evolution, the system will reach an equilibrium at $\dot{M}_\mathrm{X}=\dot{M}_\mathrm{B}<\dot{M}_\mathrm{A}$ where the ISM accretion fully accounts for the NS luminosity. 
At this stage, a disk structure could remain but it would become thicker as it is fed by the ISM accretion and $\dot{M}_\mathrm{A}$ does not decrease over time. 
If the initial mass of the fall-back disk is small, the system could also evolve to the spherical ISM accretion regime.

We assume that the fall-back disk is formed, or maintained by the ISM, and is associated with Calvera.
The accreted matter fall almost Keplerianly towards the Alfv\'en radius $r_\mathrm{A}$\citep{1979ApJ...234..296G}
\begin{eqnarray}
r_\mathrm{A}&=&\bigg(\frac{B^2R^6}{\dot{M}_\mathrm{A}\sqrt{2GM}}\bigg)^{2/7} \nonumber\\
&=& 6.18\times10^{8}B_{12}^{4/7}R_\mathrm{km}^{12/7}M_1^{-1/7}\dot{M}_{\mathrm{A},10}^{-2/7}\,\mathrm{cm},\label{eq:r_A}
\end{eqnarray}
where $B_{12}$ is the surface magnetic field in units of $10^{12}\,\mathrm{G}$, $R_\mathrm{km}$ the stellar radius in units of $\mathrm{km}$, $M_\mathrm{1}$ the stellar mass in units of $M_\mathrm{\odot}$, $\dot{M}_{\mathrm{A},10}$ is the accretion rate at $r_\mathrm{A}$ in units of $10^{10}\,\mathrm{g\,s^{-1}}$.
Matter accumulated at $r_\mathrm{A}$ will be forced to co-rotate with the NS and most of the mass would be expelled centrifugally due to the propeller effect \citep{1975A&A....39..185I}.
Consequently, the co-rotation and deflection of the matter would exert a negative torque $N$ on the star which contributes to the spin-down of the pulsar \citep{2014RAA....14...85L},

\begin{eqnarray}
N &=& 2 \dot{M_\mathrm{A}} r_\mathrm{A}^2 \Omega_\mathrm{K} (r_\mathrm{A})\bigg[1-\bigg(\frac{\Omega}{\Omega_\mathrm{K}(r_\mathrm{A})}\bigg)^\chi\bigg] \nonumber\\
&=& -I\frac{2\pi}{P^2}\dot{P}, \label{eq:brake}
\end{eqnarray}
where $\Omega_\mathrm{K}(r_\mathrm{A})=(GM/r_\mathrm{A}^3)^{1/2}$ is the Keplerian angular velocity at $r_\mathrm{A}$. The factor $\chi$ is introduced to account for the the inefficiency of the propeller effect ($0<\chi<1$). This formula reduces to the prevailing form when $\chi=1$ \citep{1999ApJ...520..276M,2000ApJ...534..373C}. 
We assume the momentum inertia of the star to be $I=MR^2/2$ and the mass-radius relation for a low-mass strangeon star can be approximated by $M=4\pi\rho R^3/3$, where the density $\rho$ is a few times the nuclear density $\rho_\mathrm{n}$ \citep{2009MNRAS.398L..31L,2013MNRAS.431.3282L,2014ChPhC..38e5101G}. We fix $\rho=2\rho_\mathrm{n}$. Combining equation (\ref{eq:r_A}) and (\ref{eq:brake}), we obtain

\begin{equation}\label{eq:pdot}
\dot{P} = 4.3 \times 10^{-12}B_{12}^{8/7}
 R_\mathrm{km}^{-17/7}P_\mathrm{s} \dot{M}_{\mathrm{A},10}^{3/7}\frac{\omega_\mathrm{s}^\chi-1}{\omega_\mathrm{s}} \,\mathrm{s\,s^{-1}},
\end{equation}
where the fastness $\omega_\mathrm{s}=\Omega/\Omega_\mathrm{K}(r_\mathrm{A})$ and $\omega_\mathrm{s}>1$ in the propeller phase. Substituting $\dot{M}_\mathrm{A}$ into $\omega_\mathrm{s}$, we see the bimodality of equation (\ref{eq:pdot}), i.e, a ``rapid'' branch of solution with $\omega_\mathrm{s}>\omega_\mathrm{c}$ and the ``slow'' branch on the opposite, where $\omega_\mathrm{c}=(2/(2-\chi))^{1/\chi}$.

 Most of the mass accreted to the the Alfv\'en radius are expelled and only a small portion of the matter are accumulated onto the strangeon star surface \citep{2017arXiv170408336R}, i.e., $\dot{M}_\mathrm{A}= \eta\dot{M}_\mathrm{X}$ with $\eta > 1$.
 $\dot{M}_\mathrm{X}$ can be inferred from the X-ray luminosity, $L_\mathrm{X}=4\pi d^2F_\mathrm{X}\approx0.1\dot{M}_\mathrm{X}c^2$. We use the flux obtained by the spectral fit $F_{\mathrm{X},0.3-10\,\mathrm{keV}}=9.0\times10^{-13}\,\mathrm{erg\,cm^{-2}\,s^{-1}}$ for calculation.
 The factor $0.1$ is the approximate energy conversion efficiency which is dominated by the gravitational potential for massive star and by the strangeonization energy release for low-mass star.

\begin{figure}
\epsscale{2.5}
\plottwo{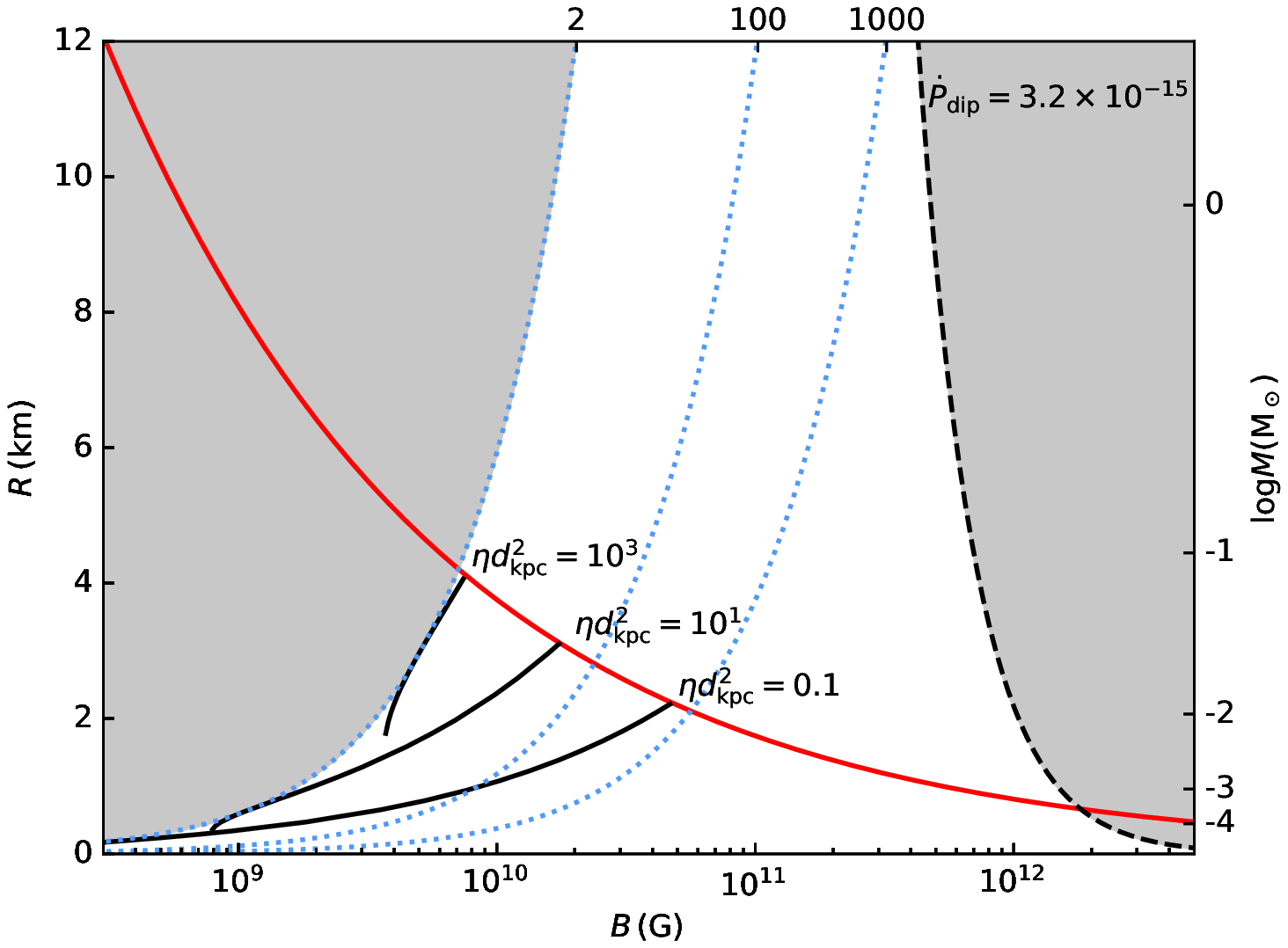}{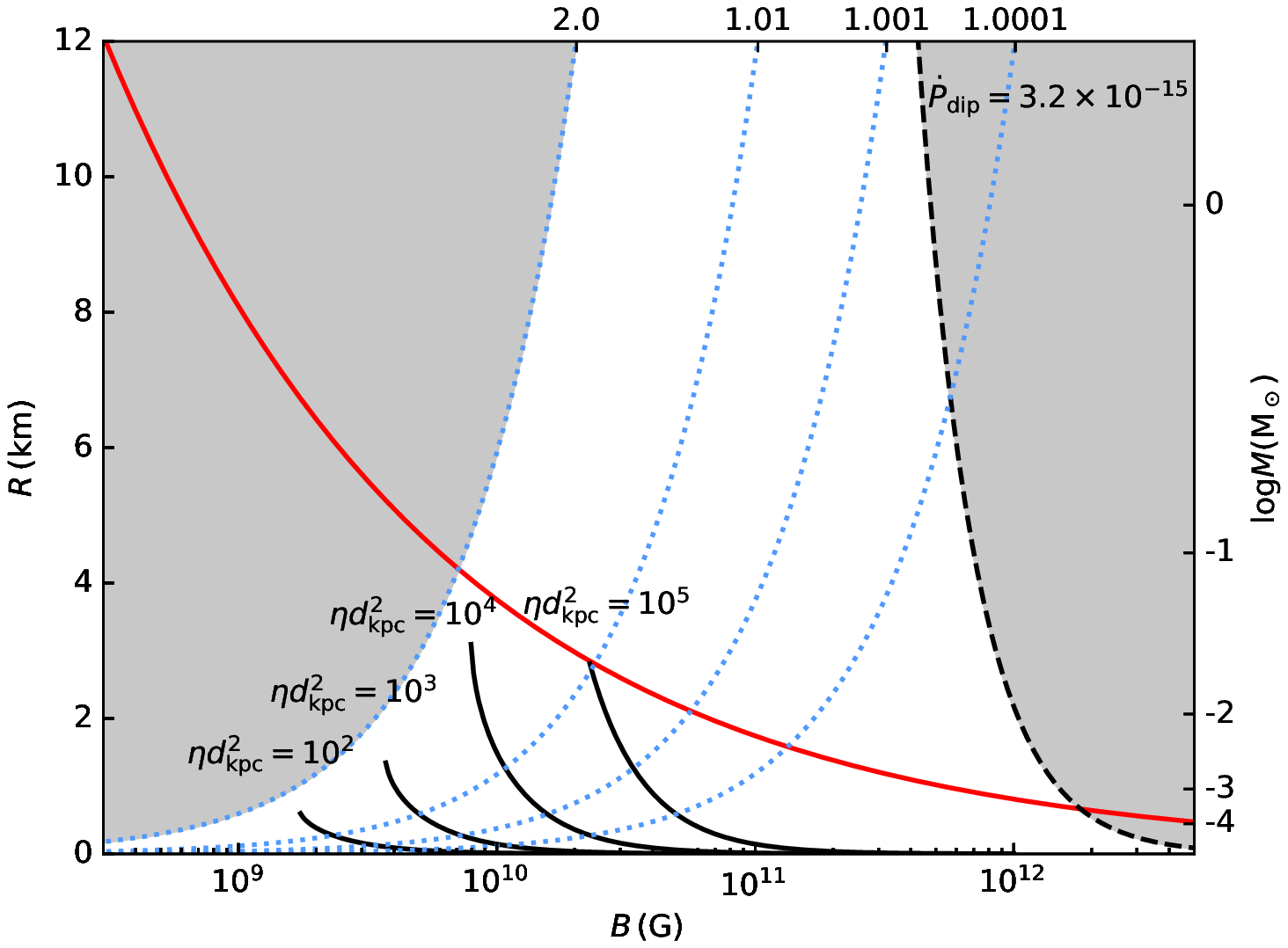}
\caption{The parameter space of $B$ and $R$ for $\chi=1$. The upper panel shows the rapid branch of the solution to equation (\ref{eq:pdot}) where $\omega_\mathrm{s}>\omega_c(\chi=1)=2$, and the lower panel shows the slow branch where $1\la\omega_\mathrm{s}<\omega_c$. The red solid curve shows the maximum permitted stellar radius of a dead strangeon star. Different choices of $\eta d_\mathrm{kpc}^2$ are plotted as black solid lines. The black dashed lines are the contours for $\dot{P}_\mathrm{dip}$ and $\dot{P}_\mathrm{dip}>3.2\times10^{-15}$ is prohibited. The blue dotted lines are contours for $\omega_\mathrm{s}$ with values tag on the top axis and the shaded area on the left corresponds to the area where equation \ref{eq:pdot} lacks solution. For most of the permitted parameter space, Calvera experiences a low surface magnetic field. \label{fig:RB_chi1}}
\end{figure}

\begin{figure}
\epsscale{2.5}
\plottwo{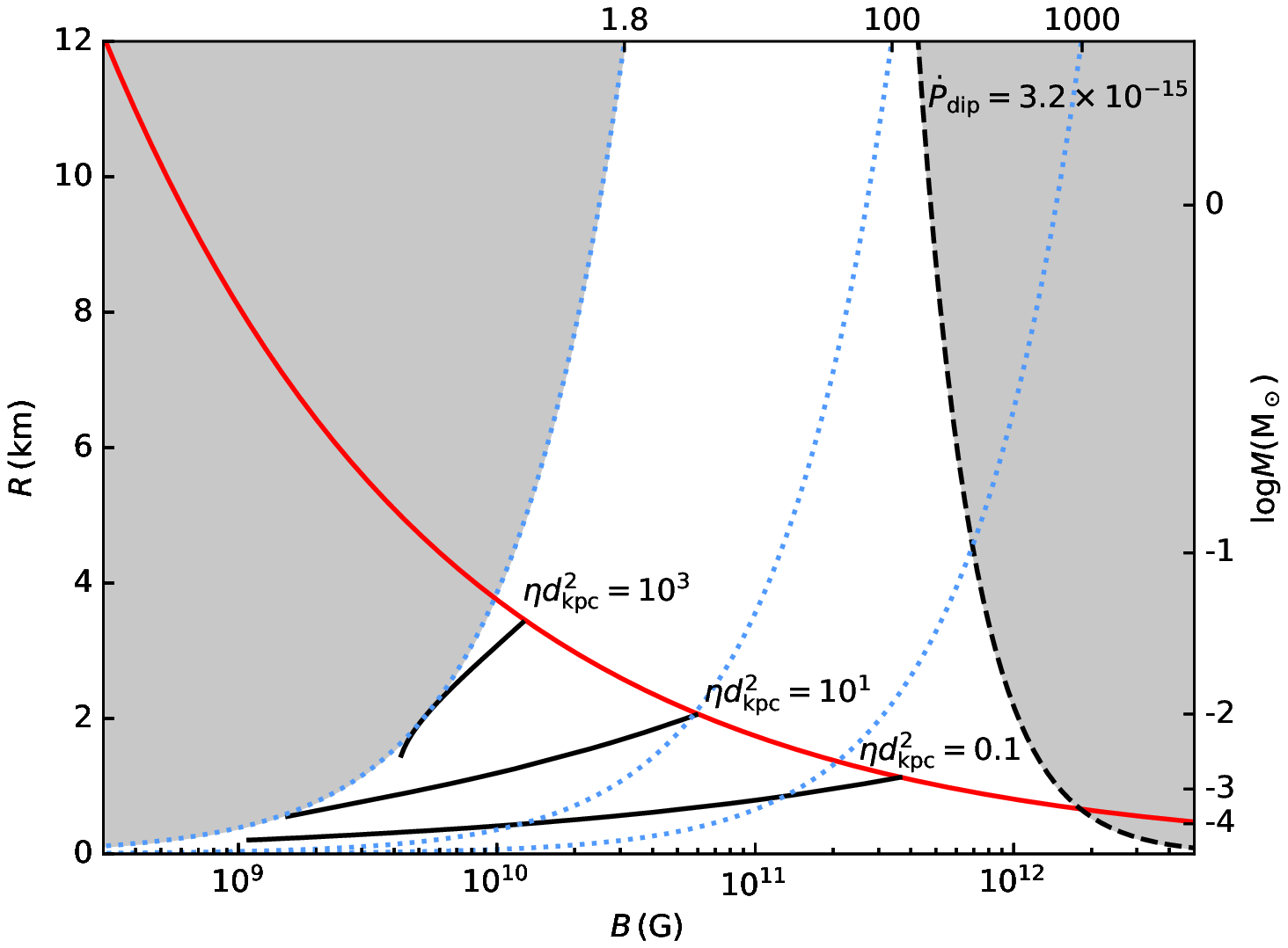}{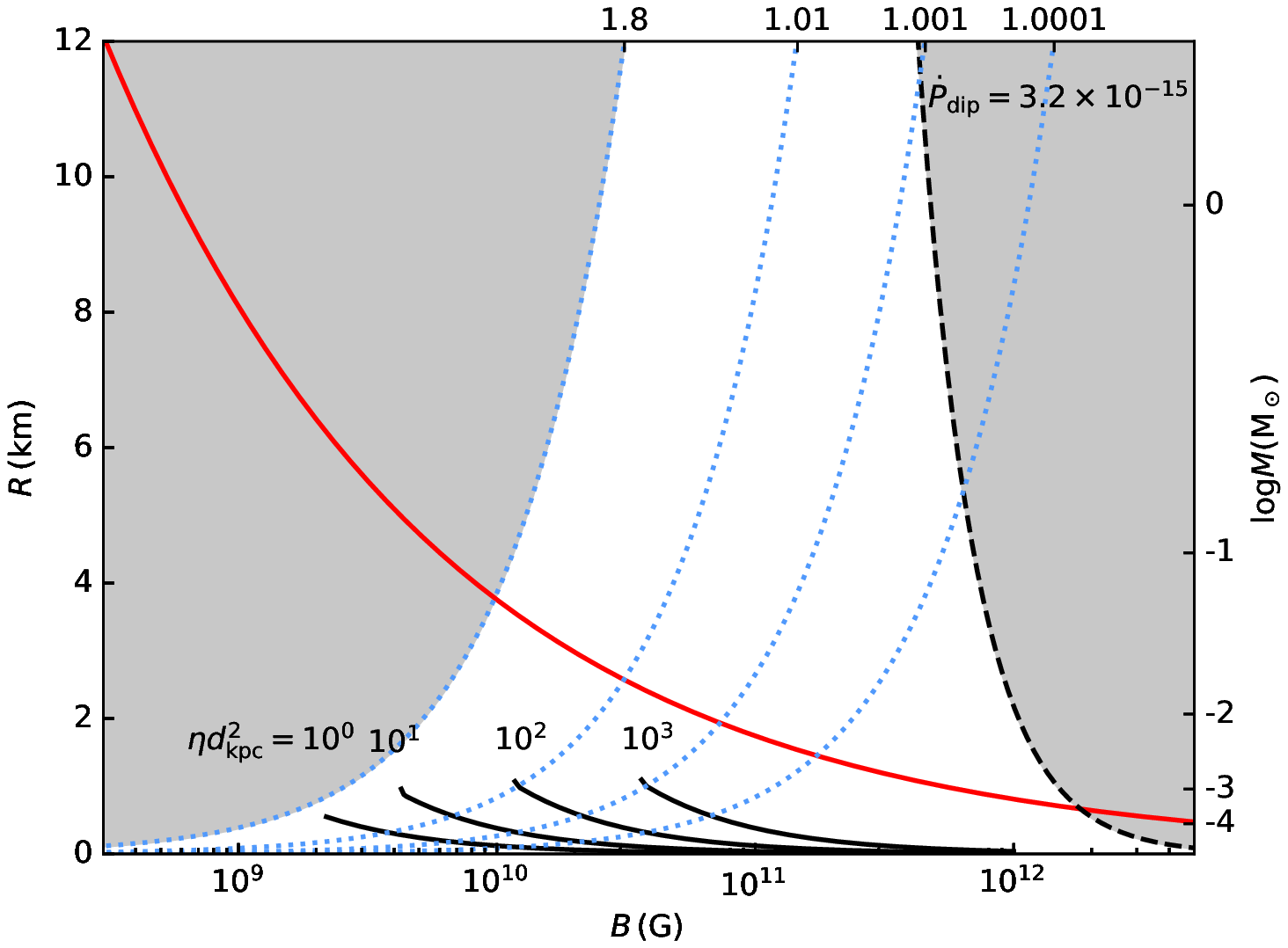}
\caption{Same with Figure \ref{fig:RB_chi1}, but for $\chi=1/2$ and $\omega_c(\chi=1/2)=1.8$. \label{fig:RB_chi2}}
\end{figure}

We present the parameter space of $B$ and $R$ in Figure \ref{fig:RB_chi1} and \ref{fig:RB_chi2}.
The red curve is the upper limit of the stellar radius beyond which the surface electric potential would exceed $\Phi_\mathrm{c}$ and the NS would be radio active (equation \ref{eq:deathline}).
Contours for $\eta d_\mathrm{kpc}^2$ are in solid black lines.
According to the definition $\eta>1$ and the speculation that Calvera can be as close as $\sim 0.3\,\mathrm{kpc}$ \citep{2015ApJ...812...61H}, $\eta d_\mathrm{kpc}^2=0.1$ can be regarded as a lower limit which yields $B\la10^{11}\,\mathrm{G}$.
The black dashed line is the contour for $\dot{P}_\mathrm{dip}$ and the shaded area on the right is prohibited.
The blue dotted lines are the $\omega_\mathrm{s}$ contour and the shaded area on the left is also banned since there is no solution to equation \ref{eq:pdot}.
The ratio of matter permeate through the Alfv\'en radius is confusing \citep{2003ApJ...588..400R,2017arXiv170408336R} and therefore left the upper limit of $\eta d_\mathrm{kpc}^2$ highly uncertain.
However, the intersection point of the blue line and the red line provides the upper limit of Calvera ($R\la4\,\mathrm{km}, M\la0.1\,M_\odot$, regardless of $\chi$), which yields a negligible redshift factor $(1-2GM/c^2R)^{-1/2}\la1.1$.
Most of the parameter space agree with a low magnetic field $B<10^{11}\,\mathrm{G}$.
The matter accreted to the polar cap diffuse across the star surface with a timescale $\tau\propto B^2$ \citep{2017ApJ...837...81W} and can cover the whole surface for a small $B$. Therefore, we assume $R=R^{\infty}$ in the following discussion.

\begin{figure}
\epsscale{2.5}
\plottwo{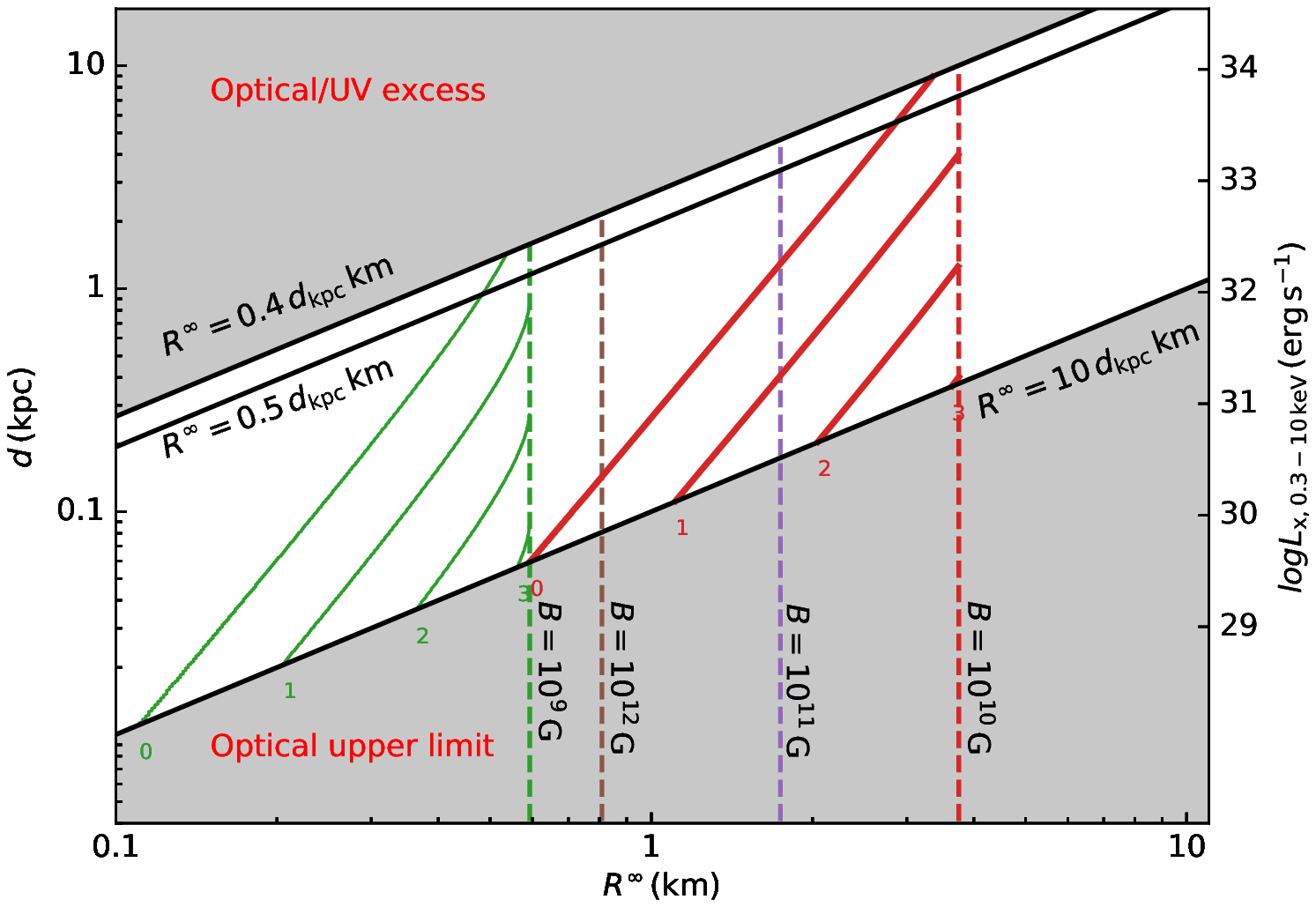}{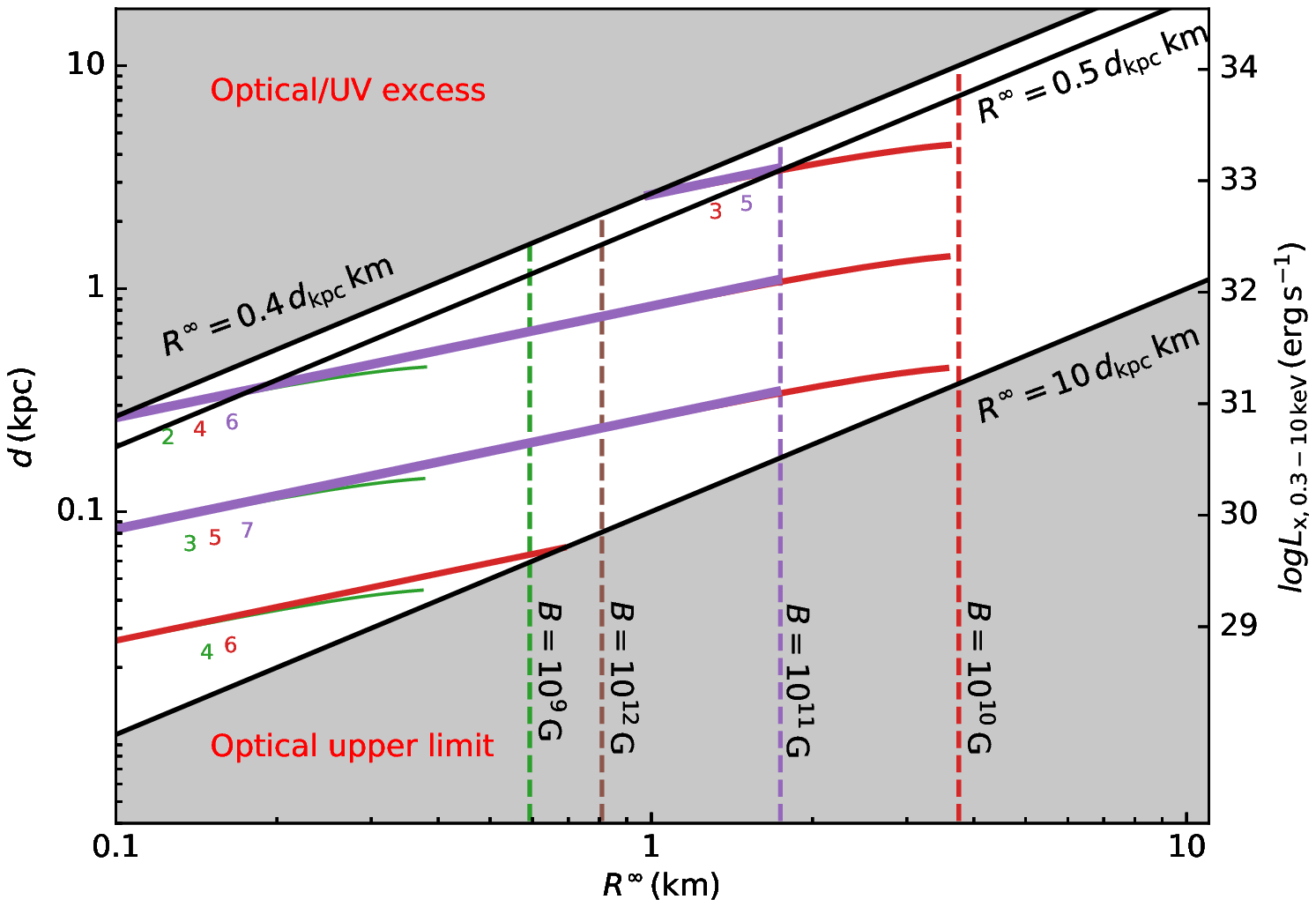}
\caption{The parameter space of radiation radius ($R^\infty$) and distance ($d$) for the rapid branch (upper panel) and the slow branch (lower panel) solution, assuming $\chi=1$. Due to the low redshift factor and the low surface magnetic field, we assume $R^{\infty}=R$. $R^\infty/d$ from model F, M1 and M2 are shown as black solid contour lines. The shaded spaces are prohibited due to the optical upper limit and optical/UV excess constraints. The colorful lines represent different combinations of $B$ and $\eta$. Lines are color-coded (also with different line thickness) by $B$ and the contours are cut off at the maximum radii (dashed lines) corresponding to $B$. The value of $\eta$ ranges from $1-10^4$ and their logarithms are tagged below each line. We regard $\eta=1$ as a lower bound, consequently,  $B\ga10^{11}$ is unlikely since the parameter space is limited. We note that these contour lines overlap with each other due to the degeneracy between $\eta$ and $B$.  (A color version of this figure is available in the online journal.)
\label{fig: r_d}}
\end{figure}

Jointing the parameters constrained by spectral modeling and the dead-pulsar criterion, we plot in Figure \ref{fig: r_d} the contours with respect to the stellar (radiation) radius $R^{\infty}$ and the distance $d$ in the $\chi=1$ case.
The upper and lower limits of the spectral normalization are plotted as black solid lines.
The colorful lines within the permitted space represent different combinations of $\eta$ and $B$ and color-coded by $B$.
The contour lines are cut off at the maximum radii (dashed lines) regarding to the value of $B$.
The maximum radius is defined by both the death line criterion and the fastness criterion, therefore a lower magnetic field might corresponds to a smaller maximum radius (see Figure \ref{fig:RB_chi1} and \ref{fig:RB_chi2}).
The logarithm of $\eta$ is tagged below each contour line.
In the rapid branch (the upper panel), the contours of $\eta$ begin at larger radii for larger $B$, i.e., a smaller maximum radius.
Therefore, beyond a certain value of $B=10^{11}\,\mathrm{G}$, no reasonable $\eta$ can satisfy the upper limit in the optical band. In the slow branch (the lower panel), the parameter space is less crowded. However, the allowed space for a high $B$ requires a large permeate fraction ($\eta>10^6$, the purple lines) that far exceeds the values obtained from simulations \citep[$\la15$,][]{2003ApJ...588..400R,2017arXiv170408336R}.
We conclude that $B \la 10^{11}\,\mathrm{G}$.

According to the observed flux of Calvera, $\dot{M}_\mathrm{X}=1.2\times10^{12}\,d^2_\mathrm{kpc}\,\mathrm{g\,s^{-1}}$. For ISM accretion, the accretion rate at the Bondi radius \citep{1952MNRAS.112..195B} is
\begin{equation}
\dot{M}_\mathrm{B}=4\pi\rho_\infty\frac{(GM)^2}{v^3},
\end{equation}
where $\rho_\infty$ is the ISM density, assumed to be $10^{-24}\rho_{24}\,\mathrm{g\,cm^{-3}}$, $v$ the speed of the star which is inferred from the proper motion measurement to be $v_\perp=286\pm110\,d_\mathrm{kpc}\,\mathrm{km\,s^{-1}}$ \citep{2015ApJ...812...61H}.
This results in $\dot{M}_\mathrm{B}=10^7\,\rho_{24}M_1^2d_\mathrm{kpc}^{-3}\,\mathrm{g\,s^{-1}}$ and
\begin{equation}
\frac{\dot{M}_\mathrm{X}}{\dot{M}_\mathrm{B}} = 10^5 \rho_{24}^{-1} M_1^{-2} d_\mathrm{kpc}^5.
\end{equation}
Even for a very small $d$ (e.g., $0.1\,\mathrm{kpc}$), the matter accretes on the stellar surface faster than the Bondi accretion and $\dot{M}_\mathrm{A}=\eta\dot{M}_\mathrm{X}\approx10^{11-15}\,\mathrm{g\,s^{-1}}$ is even higher.
The latter rate is typical for a fall-back disk of $10^5\,\mathrm{yr}$ \citep{2001ApJ...557L..61A,2009ApJ...702.1309E}, indicating that Calvera is at the early phase of the IFDA evolution, which is often expected from a pulsar with small period.

\begin{figure}
\epsscale{1.3}
\plotone{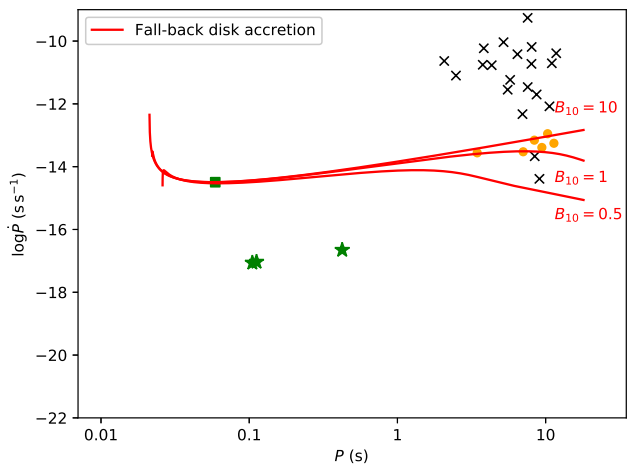}
\caption{The evolution tracks of Calvera in the IFDA model (red), with an assumption of $\chi=1$ and a stellar radius $R=2\,\mathrm{km}$. Magnetic field $B=B_{10}\times 10^{10} \mathrm{G}$. Pulsar notations are same with those in Figure \ref{ppdot}.  
We use $\dot{M_\mathrm{A}}=\dot{M}_0(t/T)^{-\alpha}$ where the time scale $T=1000\,\mathrm{s}$, $\alpha=7/6$ and $\dot{M_0}\sim10^{21-25}\,\mathrm{g\,s^{-1}}$.
\label{fig: evolution}}
\end{figure}

We illustrate the evolution of Calvera in Figure \ref{fig: evolution}, assuming $R=2\,\mathrm{km}$ and different $B$ values. 
The decrease of the accretion rate of the fall-back disk is modeled by a power-law \citep{2001ApJ...554L..63M}  $\dot{M}_\mathrm{A}=\dot{M}_0(t/T)^{-\alpha}$ (red lines). 
We set $T=1000\,\mathrm{s}$ and the current age of Calvera to be the characteristic age $3.2\times10^5\,\mathrm{yr}$ but the detailed values are not very important. $\alpha=7/6$ is adopted from \cite{2014RAA....14...85L} which is similar to that in \cite{1990ApJ...351...38C}. 

Though unlikely to be the accretion source of Calvera, we note that the ISM accretion predicts a braking index  $n=\ddot{\Omega}\Omega/\dot{\Omega}^2\approx\chi$ (assuming that the ISM feed the disk at a constant rate).
The fall-back disk model, on the other hand, predicts an infinite large braking index in the early evolution (last for $\sim 10^2\,\mathrm{yr}$, unlikely to be observable) which falls to a value $\sim 1$ during the migration. It also shows that Calvera would eventually evolve to the XDINS region within $10^7\,\mathrm{yr}$ for $B_{10}=10$, or $10^8\,\mathrm{yr}$ for $B_{10}=0.5$, and populates in a clustered area despite different $B$ values. 
Therefore, Calvera could be a progenitor of XDINS which is self-consistent since the radiation properties of the latter is also interpreted in the framework of the SSA model \citep{2017ApJ...837...81W}. In this late phase, transition from the fall-back disk accretion to ISM accretion may occur \citep{2017arXiv170503763W}.

\section{Discussion} \label{sec:disc}

\subsection{Distance}
The distance of Calvera is highly uncertain due to the lack of radio observation and optical counterpart identification. Generally, the distance estimation can be obtained through three methods: (1) Luminosity; (2) $N_\mathrm{H}$ column; and (3) proper motion measurement. Unfortunately, the first two methods are spectral-model dependent and the luminosity is even less constrained in the SSA model due to the lack of optical data. For the lower limit of $R^\infty$ in the {\it XMM-Newton} spectrum fit, we have a lower limit of $N_\mathrm{H} = (1.3\pm0.2) \times10^{20}\, \mathrm{cm^{-2}}$. This value is comparable to the galactic value $2.65\times10^{20}\,\mathrm{cm^{-2}}$ \citep{2005A&A...440..775K}, placing Calvera beyond the radius of the Local Bubble \citep{1987ARA&A..25..303C}. For a conservative estimation, we suggest $d>100\,\mathrm{pc}$ \citep{2003A&A...411..447L,2015ApJ...800...14M,2015ApJ...812...61H} as the distance lower limit.

The model independent estimation comes from the proper motion measurement. \cite{2015ApJ...812...61H} obtain the proper motion of Calvera to be $69\pm26\,\mathrm{mas\,yr^{-1}}$ corresponding to a transverse velocity $v_\perp=286\pm110\,d_\mathrm{kpc}\,\mathrm{km\,s^{-1}}$ with respect to the local standard of rest. 
Given the typical transverse velocity of XDINS to be $150-300\,\mathrm{km\,s^{-1}}$ \citep[e.g.,][]{2008AIPC..983..331K} and its high Galactic latitude, it is not likely that Calvera is a far away pulsar and $d\la1\,\mathrm{kpc}$ can be a hypothetical upper limit.

\subsection{Calvera as a CCO}
The connections between Calvera and the CCO have been hotly debated since its discovery. The non-detection of the supernova remnant \citep[within $2\arcdeg$,][]{2008ApJ...672.1137R} places Calvera as a candidate of the first orphaned CCO. In this work, we provide more evidence for this argument.

The best-fit temperature of Calvera in the context of a neutron star atmosphere result in $\la0.2\,\mathrm{keV}$ \citep{2011MNRAS.410.2428Z,2016ApJ...831..112S}---smaller than that of known CCOs which are within the range $0.3-0.7\,\mathrm{keV}$ \citep[e.g.,][also see Figure 2 in \citealt{2008ApJ...672.1137R}]{2006AA...454..543H,2001ApJ...559L.131P,2000ApJ...531L..53P,2006ApJ...653L..37P}. 
This was then attributed to the intrinsic cooling, however, the temperature of Calvera in the SSA model $\approx 0.6\,\mathrm{keV}$ readily fits in the CCO population.

The main counter-argument for Calvera being a CCO is its upper position in the $P-\dot{P}$ diagram (i.e., the high magnetic field).
\cite{2011MNRAS.414.2567H} propose that the magnetic field can be buried by prompt fall-back supernova ejecta and be recovered within $10^4\,\mathrm{yr}$.
In our accretion-braked dead strangeon star scenario, the magnetic field of Calvera is constrained at $B\la10^{11}\,\mathrm{G}$ by its flux upper limit in the optical band, which brings Calvera closer to the CCO family.
In either picture, Calvera can be interpreted as a (orphaned) CCO.
A discriminative probe would be the future measurement of the braking index. 
For an accretion braked pulsar, $n>0$ in the early phase, whereas CCO with rapid field growth would exhibit a large negative braking index \citep{2012MNRAS.425.2487V,2013ApJ...770..106B}. 
Unifying CCO and XDINS within the IFDA picture will be presented in a coming paper.

\section{Summary} \label{sec:summary}
In the framework of the strangeon star model, we conclude that Calvera is a dead low-mass strangeon star ($
\la0.1\,M_\odot$) with a small radius ($\la 4\,\mathrm{km}$) and a presumably weak magnetic field ($\la 10^{11}\,\mathrm{G}$) which is most likely braked by the fall-back disk accretion.

Nevertheless, a decisive judgment on the nature of Calvera will only come from future observations. The optical flux measurement will be crucial in determining the $d-R^{\infty}$ relation (Figure \ref{fig: r_d}). In our optical/UV excess picture, we predict the lower limit of the optical magnitude to be $g^\prime\la35$ which is challenging even for future instruments \citep[e.g., Thirty Meter Telescope (TMT),][]{2008SPIE.7012E..1AN}. However, it is possible that the optical flux is higher than the lower limit by a factor of $5-12$ \citep{2011ApJ...736..117K}, making it more accessible.
Future timing analysis can distinguish the braking mechanisms (either by accretion or by magnetic dipole radiation or by rapid magnetic field growth).
The long-term timing monitoring can be achieved with the enhanced X-ray Timing and Polarimetry \citep[eXTP,][]{2016SPIE.9905E..1QZ}; the Neutron star Interior Composition ExploreR \citep[NICER,][]{2012SPIE.8443E..13G};
the Advanced Telescope for High-ENergy Astrophysics \citep[Athena,][]{2016SPIE.9905E..2FB} and the Lynx mission \citep{2015SPIE.9601E..0JG}. If Calvera is indeed a near pulsar, a distance measurement will also benefit from the future deep optical observation or soft X-ray timing.
Although not possible at present \citep{2011MNRAS.410.2428Z}, future detection of radio and gamma-ray emission will differentiate whether Calvera is a dead pulsar.

\acknowledgments
We thank Dr. Andrey Danilenko for drawing our attention to the interesting object, Calvera, during his visiting KIAA.
We are grateful to all members in the pulsar group at Peking University and Dr. Xiangdong Li at Nanjing University for discussions.
This work is supported by the National Natural Science Foundation of China
(no. 11673002 and U1531243) and the Strategic Priority Research Program of CAS (no. XDB23010200).
\software{Xspec \citep{1996ASPC..101...17A}, CIAO \citep{2006SPIE.6270E..1VF}, SAS \citep{2014ascl.soft04004S}}
\bibliography{manuscript}
\end{document}